\documentclass[structabstract]{aa} 
\usepackage{amsmath}
\usepackage{natbib}
\usepackage{graphicx}
\usepackage{txfonts}

\newcommand{\beqy}{\begin{eqnarray}}
\newcommand{\eeqy}{\end{eqnarray}}
\newcommand{\bmlet}{\begin{subequations}}
\newcommand{\emlet}{\end{subequations}}

\def\gsimeq{\,\,\raise0.14em\hbox{$>$}\kern-0.76em\lower0.28em\hbox  {$\sim$}\,\,}  
\def\lsimeq{\,\,\raise0.14em\hbox{$<$}\kern-0.76em\lower0.28em\hbox  {$\sim$}\,\,}  

\begin{document}

   \title{Phase transitions in dense matter and the maximum mass of neutron stars}

   \subtitle{}

   \author{N.~Chamel\inst{1},
          A.~F.~Fantina\inst{1},
		  J.~M.~Pearson\inst{2},	  
	      and
          S.~Goriely\inst{1}
          }

   \institute{Institut d'Astronomie et d'Astrophysique, CP226, Universit\'e Libre de Bruxelles, B-1050 Brussels, Belgium \\
             \email{nchamel@ulb.ac.be}
        \and
         D\'ept. de Physique, Universit\'e de Montr\'eal, Montr\'eal (Qu\'ebec), H3C 3J7 Canada
         }

   \date{\today}

 
  \abstract
   {The recent precise measurement of the mass of pulsar PSR J1614$-$2230, as well 
as  observational indications of even more massive neutron stars, has revived 
the question of the composition of matter at the high densities prevailing inside 
neutron-star cores.}
    {We study the impact on the maximum possible neutron-star mass of an ``exotic"
core consisting of non-nucleonic matter. For this purpose, we study the occurrence of a 
first-order phase transition in nucleonic matter.}
   {Given the current lack of knowledge of non-nucleonic matter, we consider the stiffest 
possible equation of state subject only to the constraints of causality and thermodynamic 
stability. The case of a hadron-quark phase transition is discussed separately. The purely 
nucleonic matter is described using a set of unified equations of state that have been recently 
developed to permit a consistent treatment of both homogeneous and inhomogeneous phases.
We then compute the mass-radius relation of cold nonaccreting neutron stars with and without exotic 
cores from the Tolman-Oppenheimer-Volkoff equations.
}
   {We find that even if there is a significant softening of the equation of state associated 
with the actual transition to an exotic phase, there can still be a stiffening at higher 
densities closer to the center of the star that is sufficient to increase the maximum possible 
mass. However with quarks the maximum neutron-star mass is always reduced by assuming that the sound 
speed is limited by $c/\sqrt{3}$ as suggested by QCD calculations.
In particular, by invoking such a phase transition, it becomes possible to support 
PSR J1614$-$2230 with a nucleonic equation of state that is soft enough to be compatible with 
the kaon and pion production in heavy-ion collisions. }
   {}


   \titlerunning{Phase transitions and the maximum neutron-star mass}
   \authorrunning{N. Chamel et al.}
   \maketitle
%

\section{Introduction}
\label{intro}

Born from the catastrophic gravitational core collapse of massive stars 
(mass $M\gtrsim 8 M_\odot$)
at the end point of their evolution, neutron stars are 
among the most compact objects in the universe \citep{haen07}, with a radius 
of the order of 10 km and a mass of around 2 $M_\odot$. 
A few meters below the surface at densities above $\sim 10^4$ g.cm$^{-3}$, 
matter is so compressed that all atoms are fully ionized, and are arranged on a regular 
Coulomb lattice of nuclei, neutralized by a gas of degenerate electrons; this
is the ``outer crust". Deeper in the star, nuclei become more and more neutron 
rich, as a result of electron capture, and at a density of about 
$\sim 4\times 10^{11}$ g.cm$^{-3}$ neutrons begin to drip out of the nuclei. This 
marks the transition to the inner crust, an inhomogeneous assembly of
neutron-proton clusters and unbound neutrons, neutralized by the degenerate
electron gas \citep{pr95, lrr}. When the density reaches about $\sim 10^{14}$~g.cm$^{-3}$ 
(about half the density found at the center of heavy nuclei), the crust dissolves into 
a uniform plasma of neutrons with a small admixture of protons, neutralized by 
electrons and, at slightly higher densities, muons. 

The mass and radius of a neutron star are determined by the equation of state 
(EoS) over the full range of densities found in the star, i.e., by the 
relation between the pressure $P$ and the mass-energy density $\rho$, although the core 
will play a dominating role. This has motivated many studies of 
purely nucleonic neutron-star matter (N*M), 
i.e., a homogeneous and electrically charged neutral liquid of nucleons and leptons in 
beta equilibrium. These studies consist of simple extensions of the large 
number of many-body calculations performed since the beginning of the 1950s on 
so-called nuclear matter, consisting of just neutrons and protons that interact via 
``realistic'' forces fitted directly to experimental nucleon-nucleon phase shifts and 
to the properties of bound two- and three-nucleon systems (the Coulomb force being switched off). 
The EoS of purely nucleonic N*M has been determined in such many-body calculations up to the 
highest densities found in the most massive neutron stars. The maximum neutron-star mass obtained 
using different many-body methods and realistic forces is predicted to lie in the range between 
$\sim 1.8-2.5 M_\odot$~\citep{ls08,fuchs08} and is therefore compatible with the measured value 
$1.97\pm0.04 M_\odot$ for the mass of the pulsar PSR J1614$-$2230~\citep{dem10}. 

However, the core of massive neutron stars is likely to contain not only nucleons and leptons 
but also other particles like hyperons, meson condensates, or even deconfined quarks~\citep{page06,weber07}. 
Neutron stars with a hyperon core are sometimes referred to as ``hyperon stars'' \citep{glen}.  
Now according to Brueckner-Hartree-Fock calculations using realistic two- and three-body 
forces~\citep{vid11,bur11,sch11}, the appearance of hyperons in dense matter softens 
the EoS considerably thus lowering the maximum neutron-star mass to an almost unique value 
around $1.3-1.4 M_\odot$. On the other hand, some relativistic mean-field calculations including 
hyperons can support neutron stars that are as massive as PSR J1614$-$2230~\citep{bed11,sul12,jia12,weis12,zhao12}. 
This discrepancy can be understood 
at least partly from the fact that the maximum mass is very sensitive to the various hyperonic 
couplings, and these are determined very poorly since the limited nuclear and hypernuclear data that 
are relevant constrain the EoS only in the vicinity of the saturation density, whereas the maximum 
neutron-star mass is mostly determined by the EoS at much higher densities. Likewise, some relativistic 
mean-field models including meson condensates are able to predict the existence of massive neutron 
stars~\citep{gup12}. 

Another possibility for raising the maximum mass of neutron stars above 
$\sim 2 M_\odot$ lies in the deconfinement of quarks in the core 
\citep[such stars are generally called ``hybrid stars'', see e.g.][]{glen}. 
However, most of the calculations on which this conclusion is based are
phenomenological in the sense that they lack a direct relationship with 
quantum chromodynamics (QCD): see, e.g. \citet{alf07} and references therein. 
Even though recent perturbative QCD calculations lead to predicting  
compact stars compatible with PSR J1614$-$2230~\citep{kur10}, these calculations 
are not strictly valid for the densities prevailing in neutron stars, even in the 
most massive ones. In any case, whether the densities reached in neutron stars 
are high enough for deconfinement to occur is still an open question.

The EoS of neutron-star cores allowing for the presence of all kinds of particles 
(nucleons, leptons, hyperons, meson condensates, deconfined quarks, etc.) thus remains highly uncertain, and this 
situation is unlikely to be changed in the near future. Indeed, understanding the properties of high-density 
matter would require a consistent treatment of the various hadron species taking their internal 
structure into account based on QCD. Unfortunately, solving the equations of QCD in the nonperturbative regime prevailing 
in neutron-star cores appears as an extremely challenging problem. In comparison, the ambiguities in the EoS of 
purely nucleonic matter are much less acute, despite the divergences between different calculations at
the higher densities found in neutron-star cores.   

In this paper we do not deal with all the large uncertainties in the underlying
physics of the EoS of the ``exotic'' non-nucleonic matter that might be found 
in neutron-star cores. Rather, we determine in all generality the optimal possible increase
in the maximum neutron-star mass over what is found with purely nucleonic N*M. We only 
impose the constraints of causality and thermodynamic stability, i.e. the condition that 
at a given pressure a phase transition will occur only if the Gibbs free energy per nucleon
is lowered. For the purely nucleonic N*M, we use a set of unified EoSs 
based on the nuclear energy-density functional theory \citep{gcp10}, as 
described in Sect.\ref{nucl-eos}. In Sect.\ref{thermo} we discuss the thermodynamics of a 
possible transition to an exotic phase, and then the EoS of such a phase. For
the latter we consider first the case where the stiffness of the exotic phase
is only limited by causality, i.e., the requirement that the speed of sound
cannot exceed the speed of light, $c$. We then take account of the fact that,
according to both perturbative QCD calculations at zero temperature \citep{kur10} and 
non-pertubative lattice QCD calculations at finite temperatures \citep[see e.g.][and 
references therein]{kar07,bor10}, the speed of sound in a gas of deconfined quarks cannot exceed 
$c/\sqrt{3}$, and accordingly modify our causally limited EoS. The implications for the 
maximum neutron-star mass are discussed in Sect.\ref{NS-mass}, while in Sect.\ref{conclusion} 
we summarize our conclusions.

\section{The nucleonic equation of state} 
\label{nucl-eos}

To assess the role of a transition to a non-nucleonic
phase we must begin with a purely nucleonic EoS that has been well adapted
to the description of neutron stars whose cores are assumed to be non-exotic.
Suitable such starting points are provided by the family of three EoSs that we 
have developed to provide a unified treatment of all parts of neutron 
stars. These EoSs are based on nuclear energy-density functionals that have
all been derived from effective interactions that are generalizations of the 
conventional Skyrme forces in that they contain 
terms that depend simultaneously on momentum and density~\citep{gcp10}. 
The parameters of this form of Skyrme 
force were determined primarily by fitting measured nuclear masses, which were 
calculated with the Hartree-Fock-Bogoliubov (HFB) method. For this it was 
necessary to supplement the Skyrme forces with a microscopic contact pairing 
force, phenomenological Wigner terms, and correction terms for the spurious 
collective energy. In fitting the mass data, we simultaneously 
constrained the Skyrme force to fit the zero-temperature EoS of homogeneous 
pure neutron matter (NeuM), as determined by many-body calculations with 
realistic two- and three-nucleon forces. Actually, several such calculations 
of the EoS of NeuM have been made, and while they all agree very closely at 
nuclear and subnuclear densities, they differ in the stiffness that they 
predict at the much higher densities that can be 
encountered towards the center of neutron stars. Functional BSk19 was fitted to a soft EoS of NeuM (the one 
labeled ``UV14 plus TNI" in \citealt{wir88}, combined with the EoS of \citealt{fp81}), 
BSk21 to a very stiff EoS (the one labeled ``V18" in \citealt{ls08}), while BSk20 was 
fitted to an EoS of intermediate stiffness (the one labeled ``A18 + $\delta\,v$ + UIX$^*$" 
in \citealt{apr98}, which we abbreviate as APR). 
All three EoSs are consistent with quantum Monte Carlo calculations~\citep{gcr12}.
Even though our EoSs were not fitted to 
symmetric nuclear matter, they are all consistent with the constraint of \citet{dan02} 
deduced from heavy-ion collisions. Furthermore, the strength of the pairing force
at each point in the nucleus in question was determined so as to exactly reproduce 
realistic $^1S_0$ pairing gaps of homogeneous nuclear matter of the appropriate density and charge
asymmetry \citep{cha10}. Finally, we imposed on these forces a number of
supplementary realistic constraints, the most notable of which is the
suppression of an unphysical transition to a spin-polarized configuration, both
at zero and finite temperatures, at densities found in neutron stars and
supernova cores \citep{cgp09, cg10, gcp10}. The form of our functionals was
flexible enough for us to satisfy all these constraints and at the same 
time fit the 2149 measured masses of nuclei with $N$ and $Z \ge$ 8 given in the
2003 Atomic Mass Evaluation (AME) \citep{audi03} with an rms deviation as low as
0.58 MeV for all three models, i.e., for all three options for the high-density 
behavior of NeuM.

\begin{figure}
\resizebox{\hsize}{!}{\includegraphics{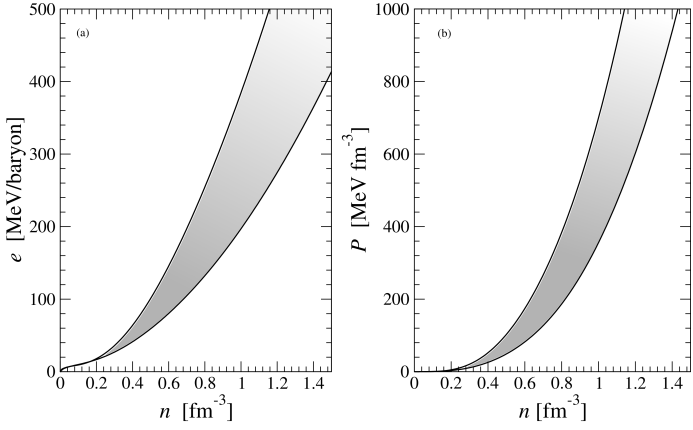}}
\caption{Left panel (a): range of energies per baryon 
(defined by $e=\mathcal{E}/n-m_n c^2$ where $m_n$ is the neutron mass) 
of nucleonic matter in beta equilibrium at zero temperature as a function of the baryon density 
for the unified Brussels-Montreal EoSs \citep{gcp10}. The shaded area reflects the different degrees
of stiffness of these EoSs: the lower limit corresponds to BSk19, the upper to BSk21. Right panel (b): corresponding 
range of pressures.}
\label{fig_eosBSk}
\end{figure}

\begin{figure}
\resizebox{\hsize}{!}{\includegraphics{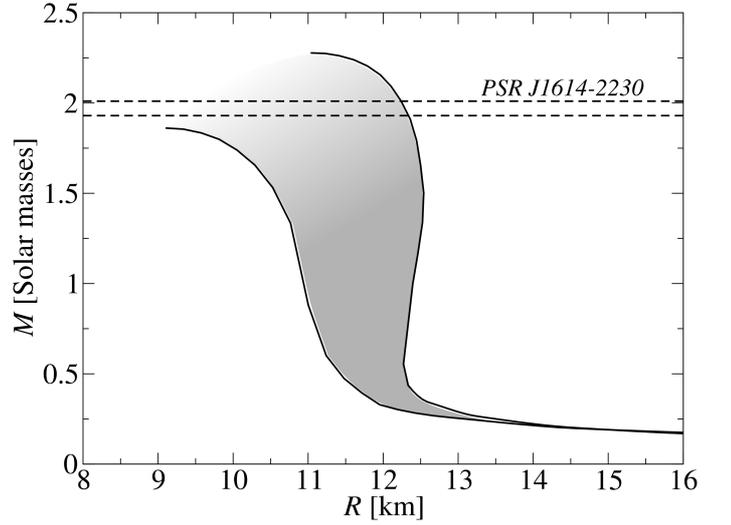}}
\caption{Range of neutron-star masses and radii for the unified 
Brussels-Montreal EoSs \citep{gcp10}. 
The shaded area reflects the different degrees of stiffness of these EoSs: 
the lower limit corresponds to BSk19, the upper to BSk21.
For comparison, we have indicated the measured mass
of PSR J1614$-$2230 including estimated errors from ~\citet{dem10}.} 
\label{fig_MR}
\end{figure}

These functionals are very well adapted to a unified treatment of all parts
of purely nucleonic neutron stars, given not only the NeuM constraints to which they have been 
subjected but also the precision fit to masses, which means that the presence 
of inhomogeneities and of protons is well represented.
We used these functionals in \citet{pgc11,pcgd12}
to calculate the properties of 
the outer and inner crusts, respectively, while in \citet{cfpg11} we 
determined the maximum possible neutron-star mass for each functional. For this
calculation we had to solve the Tolman-Oppenheimer-Volkoff (TOV) 
equations \citep{tol39,ov39} for different values of the central density,
thereby obtaining the mass $M$ of the star as a function of its radius $R$.
The total mass of a neutron star depends largely on the core properties 
\citep[although we always account for the crust using the EoS of][for 
the appropriate functionals]{pgc11,pcgd12} and thus on the EoS of homogeneous 
N*M: Fig.~\ref{fig_eosBSk} shows the range of uncertainty
spanned by our three functionals, BSk19 being the softest and BSk21 the 
stiffest. The corresponding uncertainties in the neutron-star mass for a given 
radius are shown in Fig.~\ref{fig_MR}; from this same figure we can infer that
BSk21 is stiff enough at high densities to support neutron stars as massive as 
PSR J1614$-$2230, but that functional BSk19 is too soft (BSk20 can also
support PSR J1614$-$2230). It is worth noting that our functionals are 
consistent with the radius constraints of \citet{ste10} inferred from observations of X-ray 
bursters and low-mass X-ray binaries.

Since the softest of our Skyrme functionals that is compatible with the measured 
mass of PSR J1614$-$2230 is BSk20 \citep{cfpg11}, the question arises as to whether 
it is not too stiff to be compatible with the analysis of $K^+$ production 
\citep{fuchs01,stu01,har06} and $\pi^-/\pi^+$ production ratio \citep{xiao09} that 
have been measured in heavy-ion collisions. In particular, the former analysis suggests
that the EoS of symmetric nuclear matter is much softer than what is obtained with BSk20 
over the range $2n_0 \lesssim n \lesssim 3n_0$, whereas the latter analysis concludes that over 
the range $2n_0 \lesssim n \lesssim 3.5n_0$ the symmetry energy rises significantly 
less steeply than predicted for the APR EoS, on which BSk20 is based.
These results, taken at face value, 
discriminate against both BSk20 and BSk21, and favor BSk19. Various exotic
mechanisms, such as a ``fifth force" \citep{wen09} or variations of the
gravitational constant \citep{wen12}, have been proposed to simultaneously account 
 for both this result and the existence of PSR J1614$-$2230,
but we shall see that the much more prosaic explanation of a transition to
a non-nucleonic phase may suffice.

\section{Transition to non-nucleonic phase in dense matter} 
\label{thermo}

\subsection{Thermodynamic equilibrium}

In view of the current
uncertainties about the composition of matter at supernuclear densities, we 
simply assume that above a baryon density $n_{\rm N}$, nucleonic 
matter undergoes a first-order phase transition to some unknown exotic phase 
with a baryon density $n_{\rm X}>n_{\rm N}$. 
The exotic phase could consist of a pion condensate, a kaon condensate, 
hyperonic matter, or deconfined quarks~\citep[see e.g.][]{glen,haen07}.
In the region of coexisting 
phases, $n_{\rm N}\leq n\leq n_{\rm X}$, thermodynamic equilibrium requires the
constancy of the pressure $P$ and baryon chemical potential $\mu$~:
\beqy
\label{1}
P_{\rm exo}(n)=P_{\rm nuc}(n_{\rm N}), \ \mu_{\rm exo}(n)=
\mu_{\rm nuc}(n_{\rm N}) \, ,
\eeqy
where the subscripts ``exo'' and  ``nuc'' are used to denote the exotic and 
nuclear-matter EoS, respectively. 
As a result, using the general expression
\beqy
\label{1A}
\mathcal{E} = n\mu - P   \quad ,
\eeqy
valid at the zero temperature that we assume throughout this paper,
we see that the energy density of the coexisting phases varies linearly with 
the baryon density in the range $n_{\rm N}\leq n\leq n_{\rm X}$ according to
\beqy
\label{2}
\mathcal{E}_{\rm exo}(n)= 
n \mu_{\rm nuc}(n_{\rm N}) - P_{\rm nuc}(n_{\rm N})\, .
\eeqy
At the density $n_{\rm X}$, which we always suppose to be lower than the 
central density $n_{\rm cen}$ of the star, the phase transition has reduced the
pressure from $P_{\rm nuc}(n_{\rm X})$ to $P_{\rm nuc}(n_{\rm N})$.

In this simple picture of a first-order phase transition, the two phases cannot coexist in 
the star because the densest phase will sink below the other. In hydrostatic equilibrium, 
the two phases would thus be spatially separated with the density varying discontinuously 
at the interface.
In reality, the two distinct phases will generally rearrange themselves by forming 
a mixed phase in a finite region of the star \citep{tats11} unless the surface tension 
and the Coulomb interaction are sufficiently strong, as could be the case
for the hadron-quark phase transition \citep[see e.g.][]{hps93,alf01,endo06}. 
The presence of a mixed phase leads to a smooth transition with the pressure increasing 
monotonically with the density instead of 
remaining constant. This situation generally increases the maximum neutron-star mass 
\citep[see e.g.][section 9.3]{glen}. However, given the uncertainties pertaining 
to the EoS of the mixed phase, here we suppose the least favorable case of a first-order 
transition considered in Eqs.~(\ref{1}) and~(\ref{2}) which will provide a lower bound on 
the maximum mass \citep{rhru74}. 

\subsection{Equation of state for exotic matter}

{\it Causal limit of EoS.} Instead of considering specific models of 
non-nucleonic matter, we suppose
that the EoS of the exotic phase is the stiffest possible, being only constrained 
by Le Chatelier's principle 
\beqy
\frac{dP}{d\mathcal{E}}\geq 0\, 
\eeqy
and the causality requirement 
\beqy
\frac{dP}{d\mathcal{E}}\leq 1\, .
\eeqy
Since the maximally compact stars are obtained for $dP/d\mathcal{E}=1$ \citep{rhru74}, 
we will assume that the pressure of the exotic phase at densities $n>n_{\rm X}$ can be expressed as
\beqy
\label{3}
P_{\rm exo}(n)=\mathcal{E}_{\rm exo}(n)-\mathcal{E}_{\rm exo}(n_{\rm X}) + P_{\rm nuc}(n_{\rm N})\, ,
\eeqy
recalling that $P_{\rm nuc}(n_{\rm N})=P_{\rm exo}(n_{\rm X})$. After integrating the 
pressure by using Eq.~(\ref{1A}) with $\mu = d\mathcal{E}/dn$, we find for the 
energy density of the exotic phase
\beqy
\label{4}
\mathcal{E}_{\rm exo}(n)=\frac{n_{\rm X} \mu_{\rm nuc}(n_{\rm N})}{2}\left[1+
\left(\frac{n}{n_{\rm X}}\right)^{2}\right]-P_{\rm nuc}(n_{\rm N})\, .
\eeqy
Using Eqs.~(\ref{2}), (\ref{3}), and (\ref{4}), the baryon chemical potential in the exotic phase can be expressed as 
\beqy
\label{4b}
\mu_{\rm exo}(P)=\mu_{\rm nuc}(n_{\rm N})\sqrt{1+\frac{2(P-P_{\rm nuc}(n_{\rm N}))}{n_{\rm X}\mu_{\rm nuc}(n_{\rm N})}}\, .
\eeqy
From both Eqs. (\ref{4}) and (\ref{4b}) it is seen that the EoS of the exotic 
phase is completely determined by the parameters
$n_{\rm N}$ and $n_{\rm X}$, and for any given pair of values of these parameters  we have 
to determine the maximum possible neutron-star mass. But in this very general
study we have a high degree of freedom in choosing the values of these 
parameters, and we do so
in such a way as to obtain a {\it maximum maximorum} in the neutron-star mass,
subject to certain physical constraints that we now discuss.

Dealing first with $n_{\rm N}$, we find that increasing values of this parameter
lead to decreasing values of the maximum mass, as is to be expected intuitively. We therefore
consider the lowest possible density $n_{\rm N}$ consistent with nuclear data. 
In particular, we set $n_{\rm N} = 0.2$ fm$^{-3}$, which is slightly higher than
the highest density found in nuclei, as 
predicted by HFB calculations on more than 8000 nuclei\footnote{http://www.astro.ulb.ac.be/bruslib}.
Since much higher densities can be reached in heavy-ion collisions, the lack of
any evidence of phase transitions might suggest that 
$n_{\rm N}\gg 0.2$ fm$^{-3}$. In fact, this is not necessarily the case because
the conditions prevailing in neutron-star interiors are very different from 
those encountered in heavy-ion collisions. We do not discuss here the 
possibility of strange stars for which the exotic 
phase (strange-quark matter in this case) would be present in the entire star. 
At densities $n<n_{\rm N}$, we use the EoS of \citet{pgc11,pcgd12} for the 
outer and inner parts of the crust and the EoS of \citet{gcp10} for the 
purely nucleonic part of the core.

\begin{figure}
\resizebox{\hsize}{!}{\includegraphics{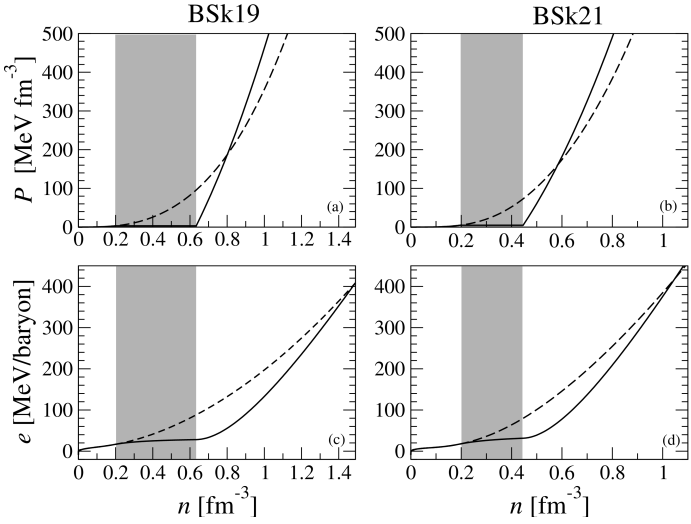}}
\caption{Upper panels: pressure as a function of the baryon number density (up to the 
highest central density $n_{\rm cen}$ found in the most massive neutron stars) for the 
softest (BSk19, panel a) and the stiffest (BSk21, panel b) of our nucleonic EoSs (dashed lines) and for 
the corresponding EoSs with a causally limited phase transition (solid lines). 
Lower panels: energy per baryon (defined by $e=\mathcal{E}/n-m_n c^2$ where $m_n$ is the neutron
mass) as a function of the baryon density for the 
softest (BSk19, panel c) and the stiffest (BSk21, panel d) of our nucleonic EoSs (dashed lines) and for 
the corresponding EoSs with a causally limited phase transition (solid lines). The shaded areas indicate the 
region of phase coexistence at densities $n_{\rm N}\leq n\leq n_{\rm X}$. 
} 
\label{fig1}
\end{figure}

\begin{figure}
\resizebox{\hsize}{!}{\includegraphics{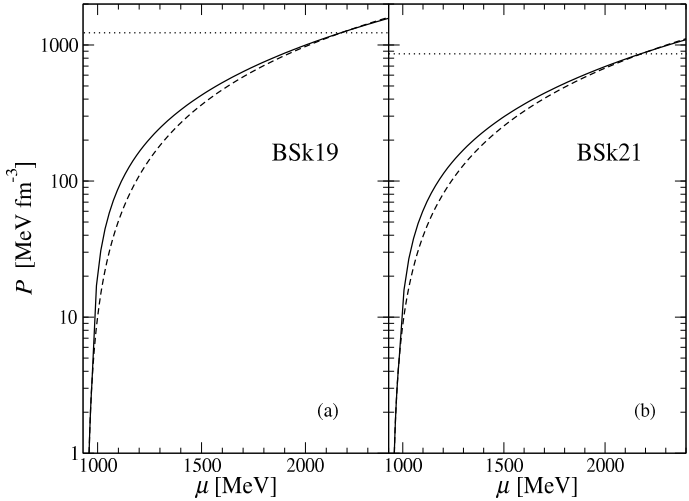}}
\caption{Pressure as a function of the baryon chemical potential for the 
softest (BSk19, panel a) and the stiffest (BSk21, panel b) of our nucleonic EoSs (dashed lines) and for 
the corresponding EoSs with a causally-limited phase transition (solid lines). 
The horizontal dotted line indicates the central pressure in the most massive neutron stars. 
} 
\label{fig1b}
\end{figure}

As for $n_{\rm X}$, we note first that the pressure of the exotic phase remains lower
than that of the nucleonic phase from the baryon density $n_{\rm N}$ up to some
density $n_{\rm P}>n_{\rm X}$, but that it is higher thereafter, as shown in 
Fig.~\ref{fig1}. Therefore the impact of a phase transition will be to increase the 
maximum neutron-star mass provided the central density $n_{\rm cen}$ is substantially 
higher than $n_{\rm P}$. However, $n_{\rm P}$ cannot be freely adjusted because of 
the requirement that the exotic phase should be energetically favored. More precisely, 
at a given pressure $P$, the equilibrium phase is found by minimizing the Gibbs free 
energy per baryon, which coincides with the baryon chemical potential. 
Now the baryon chemical potential of the exotic phase will rise more steeply than
that of the nucleonic phase, and at some pressure $P_{\rm C}$ the two phases will 
have the same baryon chemical potentials, $\mu_{\rm exo}(P_{\rm C})=\mu_{\rm nuc}(P_{\rm C})$.
This will occur when the density of the exotic phase reaches the value $n_{\rm C}$, 
such that $P_{\rm exo}(n_{\rm C})=P_{\rm C}$. For pressures $P > P_{\rm C}$, or equivalently 
for densities $n>n_{\rm C}$, the ground state of matter will once again be purely nucleonic. 
In order to exclude this unlikely possibility, we must have $P_{\rm C}$ 
higher than the central pressure $P_{\rm cen}$ (or equivalently $n_{\rm C}$
higher than $n_{\rm cen}$) in the most massive neutron stars.
Fixing $P_{\rm C}$ then completely determines the density $n_{\rm X}$, which
is given by 
\beqy
\label{5}
n_{\rm X}= \frac{2 (P_{\rm C}-P_{\rm nuc}(n_{\rm N}))}
{\mu_{\rm nuc}(n_{\rm N})}
\Biggl[\left(\frac{\mu_{\rm nuc}(P_{\rm C})}{\mu_{\rm nuc}(n_{\rm N})}\right)^2-1\Biggr]^{-1}\, ,
\eeqy
where we have used Eq.~(\ref{4b}). 
At the same time, we confirmed the intuitively plausible result that the {\it lower} $P_{\rm C}$ is, the 
greater the maximum possible mass, and accordingly we arranged for $P_{\rm C}$ to be very close to $P_{\rm cen}$. 
Now the maximum central pressure in a neutron star is approximately given by \citep{lp10} 
\beqy
\label{5b}
P_{\rm cen}\approx 2.034\, \mathcal{E}_{\rm exo}(n_{\rm X})=2.034\, (n_{\rm X} \mu_{\rm nuc}(n_{\rm N})-P_{\rm nuc}(n_{\rm N}))\, ,
\eeqy
where we have used Eq.~(\ref{4}). The optimum value of $n_{\rm X}$ was therefore obtained by solving 
Eqs.~(\ref{5}) and (\ref{5b}) with $P_{\rm cen}=P_{\rm C}$. However, since Eq.~(\ref{5b}) is only approximately valid, 
we subsequently refined the value of $P_{\rm C}$ by solving the TOV equations numerically in order to 
obtain a more accurate estimate of $P_{\rm cen}$. With $P_{\rm C}$ and $n_{\rm X}$ determined, the density $n_{\rm C}$ 
(for which $P_{\rm exo}(n_{\rm C})=P_{\rm C}$) can be obtained from Eqs.~(\ref{3}) and (\ref{4}). 
The corresponding EoSs are as shown in Figs.~\ref{fig1} and \ref{fig1b}. In particular, we note that at a given pressure 
the phase transition leads to a lowering of the baryon chemical potential. 

The density $n_{\rm P}$ is completely determined by the nucleonic EoS, $n_{\rm N}$ and 
$n_{\rm C}$, as follows. We first note that for the EoS given by Eq.~(\ref{3})
we have $\mu=d\mathcal{E}/dn=dP/dn$. Differentiating Eq.~(\ref{1A}) thus leads to
\beqy
\frac{d\mu}{\mu} = \frac{dn}{n}\, ,
\eeqy
which after integration using Eq.~(\ref{1}), yields
\beqy
\frac{\mu_{\rm exo}(n_{\rm P})}{n_{\rm P}}=\frac{\mu_{\rm exo}(n_{\rm X})}{n_{\rm X}}=\frac{\mu_{\rm nuc}(n_{\rm N})}{n_{\rm X}}\, .
\eeqy

{\it Case of quark matter}. The foregoing analysis may be
invalid and the limiting EoS much softer if quark deconfinement takes place in 
the core of a neutron star. According to both perturbative QCD calculations at zero temperature \citep{kur10} and 
nonperturbative lattice QCD calculations at finite temperatures \citep{kar07,bor10}, the speed of sound in quark 
matter is limited by $c/\sqrt{3}$. 
Assuming that this result remains valid in the interior of neutron stars, the
stiffest possible EoS is given by
\beqy\label{12}
P_{\rm quark}(n)=\frac{1}{3}\left[\mathcal{E}_{\rm quark}(n)-\mathcal{E}_{\rm quark}(n_{\rm X})\right] + P_{\rm nuc}(n_{\rm N})\, .
\eeqy
After integrating the pressure, we find for the energy density
\beqy\label{13}
\mathcal{E}_{\rm quark}(n)=\frac{3}{4} n_{\rm X} \mu_{\rm nuc}(n_{\rm N})\left[\frac{1}{3}+\left(\frac{n}{n_{\rm X}}\right)^{4/3}\right]-P_{\rm nuc}(n_{\rm N})\, ,
\eeqy
instead of Eq.~(\ref{4}). This equation of state ressembles that obtained 
with the simplest MIT bag model, which has been commonly used for describing 
the interior of compact stars \citep[see e.g.][]{glen,haen07}. In this model, 
quarks are treated as massless and noninteracting particles confined inside a 
``bag''. The pressure of the quarks is then given by
\beqy\label{14}
P_{\rm quark}(n)=\frac{1}{3}\biggl(\mathcal{E}_{\rm quark}(n)-4 B\biggr)\, ,
\eeqy
where $B$ is the bag pressure. Comparing Eq.(\ref{12}) and (\ref{14}) yields the effective bag pressure
\beqy
B=\frac{1}{4} n_{\rm X} \mu_{\rm nuc}(n_{\rm N})-P_{\rm nuc}(n_{\rm N})\, .
\eeqy
As a matter of fact, Eq.(14) is also a fairly good approximation of more 
realistic quark matter EoSs \citep{zdu00,dor00}. Using Eqs.~(\ref{12}) and (\ref{13}), 
the chemical potential is given by 
\beqy
\mu_{\rm quark}(P)=\mu_{\rm nuc}(n_{\rm N})\Biggl[\frac{4(P-P_{\rm nuc}(n_{\rm N}))}{\mu_{\rm nuc}(n_{\rm N}) n_{\rm X}}+1\Biggr]^{1/4}\, .
\eeqy

\begin{figure}
\resizebox{\hsize}{!}{\includegraphics{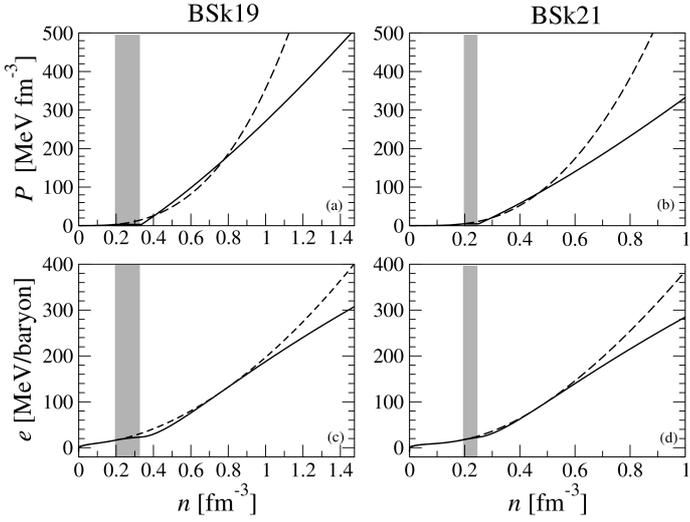}}
\caption{Upper panels: pressure as a function of the baryon number density (up to the 
highest central density $n_{\rm cen}$ found in the most massive neutron stars) for the 
softest (BSk19, panel a) and the stiffest (BSk21, panel b) of our nucleonic EoSs (dashed lines) and for 
the corresponding EoSs with a quark phase transition (solid lines). 
Lower panels: energy per baryon (defined by $e=\mathcal{E}/n-m_n c^2$ where $m_n$ is the 
neutron
mass) as a function of the baryon density for the 
softest (BSk19, panel c) and the stiffest (BSk21, panel d) of our nucleonic EoSs (dashed lines) and for 
the corresponding EoSs with a quark phase transition (solid lines). The shaded areas indicate the 
region of phase coexistence at densities $n_{\rm N}\leq n\leq n_{\rm X}$. } 
\label{fig2}
\end{figure}

\begin{figure}
\resizebox{\hsize}{!}{\includegraphics{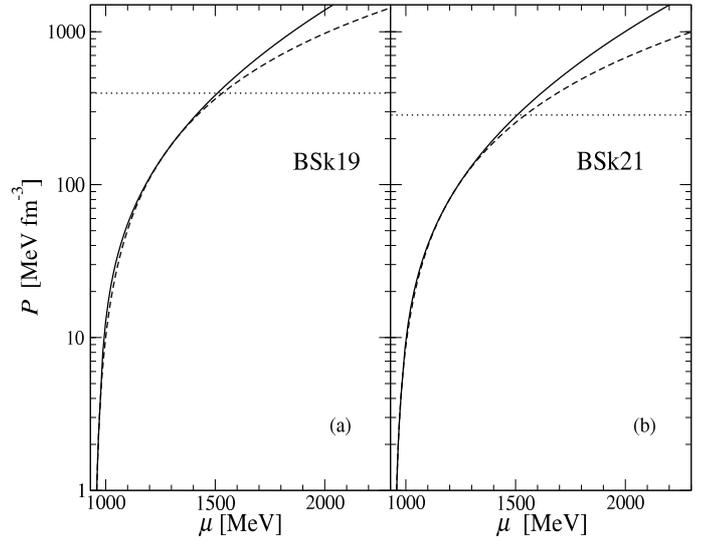}}
\caption{Pressure as a function of the baryon chemical potential for the 
softest (BSk19, panel a) and the stiffest (BSk21, panel b) of our nucleonic EoSs (dashed lines) and for 
the corresponding EoSs with a quark phase transition (solid lines). 
The horizontal dotted line indicates the central pressure in the most massive neutron stars. 
} 
\label{fig2b}
\end{figure}

Again, the EoS of the quark phase is completely determined by the parameters 
$n_{\rm N}$ and $n_{\rm X}$, whose values must be specified. We choose the former as 
before, remarking that according to percolation simulations \citep{mag03} the density $n_{\rm N}$
at which the hadron-quark phase transition begins could be very close to the saturation density.
The density $n_{\rm X}$ is still determined by maximizing the neutron-star 
mass, subject to the constraint that there be no reconversion of the exotic 
phase into nucleonic matter at the highest densities prevailing in the most 
massive neutron stars. However, the situation is now more complicated than
in the previous case, where the stiffness of the EoS is limited only by 
causality. The EoS for quark matter is shown in Figs.~\ref{fig2} and \ref{fig2b}, 
where it will be seen that it is possible to have equality of both the chemical potentials and the 
pressures of the two phases at the same density $n_{\rm C}$~:
\beqy
P_{\rm quark}(n_{\rm C})=P_{\rm nuc}(n_{\rm C})\, , \ \ \mu_{\rm quark}(n_{\rm C})=\mu_{\rm nuc}(n_{\rm C})\, .
\eeqy 
It follows from Eq.~(\ref{1A}) that the energy densities of the two phases are also equal 
\beqy
\mathcal{E}_{\rm quark}(n_{\rm C})=\mathcal{E}_{\rm nuc}(n_{\rm C})\, .
\eeqy
We found numerically 
that this situation corresponds to the maximum possible neutron-star mass. It 
particularly needs to be emphasized that while $n_{\rm C}$ is now less than 
$n_{\rm cen}$, the exotic phase is not reconverted into nucleonic matter: although 
it is on the point of doing so at $n_{\rm C}$, the EoS of quark matter immediately
softens as $n$ increases beyond $n_{\rm C}$. In any case, we see that while the 
pressure in the quark phase is higher than in the nucleonic phase at relatively
low densities, the reverse is the case at the higher densities relevant to the
maximum neutron-star mass that can be supported. We may thus anticipate that
quark deconfinement will reduce the maximum neutron-star mass, assuming that 
the speed of sound is limited by $c/\sqrt{3}$. 

Concerning the low-density phase transition from nucleonic to quark matter,
we see on comparing Figs.~\ref{fig1} and~\ref{fig2} that the density range 
$n_{\rm X} - n_{\rm N}$ over which the two phases coexist is much narrower
than in the case of the transition to the causally limited EoS.

The effective bag constants associated with the nucleonic EoSs BSk19, BSk20 
and BSk21 (respectively 78.6, 65.5, and 56.7 MeV fm$^{-3}$) lie in the range of 
values that have been generally adopted for studies of hybrid stars~\citep[see e.g.][]{haen07}. 

\section{Neutron-star maximum mass} 
\label{NS-mass}

\begin{figure}
\resizebox{\hsize}{!}{\includegraphics{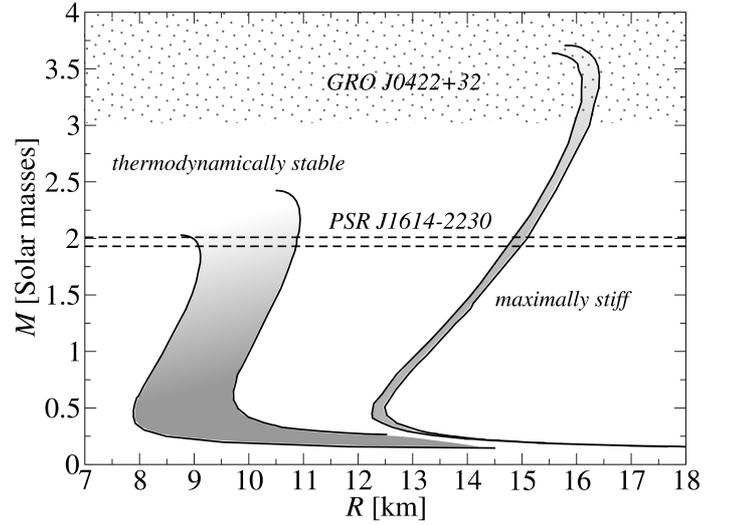}}
\caption{Range of masses and radii of neutron stars with no quark cores (shaded areas), 
for maximally stiff EoSs and for EoSs satisfying thermodynamic stability. For comparison, we  
indicate the measured mass of PSR J1614$-$2230 including estimated errors from ~\citet{dem10}. The dotted area delimits 
the estimated range of masses of the soft X-ray transient GRO~J0422$+$32 from \citet{gel03}. 
} 
\label{fig_MR_exo}
\end{figure}

\begin{figure}
\resizebox{\hsize}{!}{\includegraphics{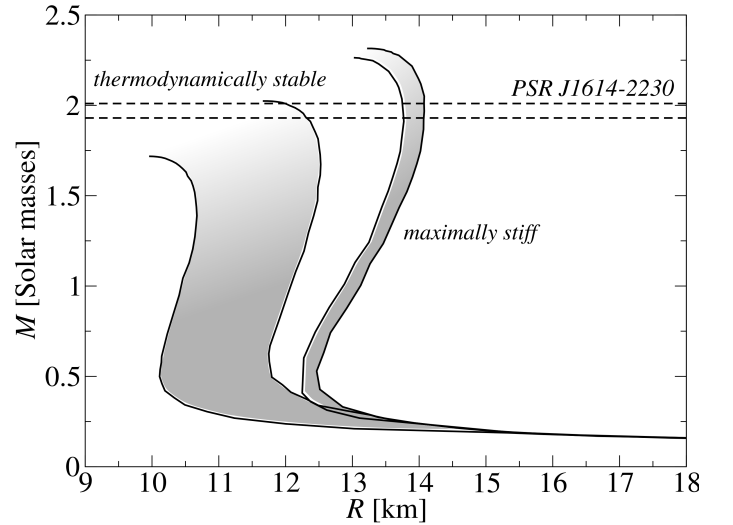}}
\caption{Range of masses and radii of neutron stars with quark cores (shaded areas), 
for maximally stiff quark EoSs and for EoSs satisfying thermodynamic stability. For comparison, we  
indicate the measured mass of PSR J1614$-$2230 including estimated errors from ~\citet{dem10}. 
} 
\label{fig_MR_qk}
\end{figure}

For each of our three unified EoSs we have solved the TOV equations that describe the 
global structure of spherical nonrotating neutron stars \citep{tol39,ov39}. As shown 
in a previous paper \citep{cfpg11}, the impact of the rotation on the maximum neutron star 
mass is negligibly small for stars having rotation periods comparable to that of PSR J1614$-$2230. 
For simplicity, we have therefore ignored rotation. The resulting mass-radius relations are plotted 
in Figs.~\ref{fig_MR_exo} and \ref{fig_MR_qk}. 

Neutron stars with central densities $n_{\rm cen}\gtrsim n_{\rm X}$ may be unstable with respect to 
radial oscillations due to the strong softening accompanying the phase transition in dense matter. 
Unstable configurations, which are characterized by the inequality \citep[see e.g.][]{haen07}
\beqy
\frac{dM}{d\mathcal{E}_{cen}} < 0 \, ,
\eeqy
are indicated in Tables~\ref{tab0} and \ref{tab} for each of our three functionals. For neutron stars 
with quark cores, no instabilities were found for BSk20 and BSk21, and unstable configurations for BSk19
were found to be restricted to a very narrow range of masses and radii. 

\begin{table}
\centering
\caption{Stability of nonrotating neutron stars with exotic cores: range of central pressures of stars
that are unstable with respect to radial oscillations, corresponding range of masses and radii.}
\label{tab0}
\vspace{.5cm}
\begin{tabular}{|c|ccc|}
\hline
Force & $P_{\rm cen}$ (MeV fm$^{-3}$) & $M/M_{\odot}$  & $R$ (km) \\
\hline
BSk19 & $3.32-11.0$ & $0.145-0.188$  & $14.5-15.5$ \\ 
BSk20 & $4.17-9.28$ & $0.207-0.233$  & $13.1-13.9$ \\ 
BSk21 & $4.87-9.49$ & $0.265-0.290$  & $12.6-13.2$ \\ 
\hline
\end{tabular}
\end{table}

\begin{table}
\centering
\caption{Same as Table~\ref{tab0} for neutron stars with quark cores.}
\label{tab}
\vspace{.5cm}
\begin{tabular}{|c|ccc|}
\hline
Force & $P_{\rm cen}$ (MeV fm$^{-3}$) & $M/M_{\odot}$  & $R$ (km) \\
\hline
BSk19 & $3.32-11.0$ & $0.186-0.188$ & $15.32-15.33$ \\ 
\hline
\end{tabular}
\end{table}

%

The numerical values of the maximum neutron-star masses $M_{\rm max}$ 
are indicated in Tables~\ref{tab1} and \ref{tab2} for each of our three functionals. We also 
show in these tables the corresponding radius, the highest baryonic density 
$n_{\rm N}$ of the nucleonic phase, the lowest baryonic density $n_{\rm X}$ of the exotic 
phase, and the central baryonic density $n_{\rm cen}$. 

In Table~\ref{tab3} we summarize the results of \citet{cfpg11} for the
case of no exotic phase. Comparing with Table~\ref{tab1}, we see that the
result of allowing a transition to an exotic phase whose EoS is limited
only by causality is always to increase the maximum possible neutron-star mass.
Of particular interest is the case of functional BSk19, which can now
support pulsar PSR J1614$-$2230. Thus this functional allows us to reconcile 
the existence of this pulsar with the $K^+$ production \citep{fuchs01,stu01,har06}
and the $\pi^-/\pi^+$ production ratio measured in heavy-ion collisions \citep{xiao09}, 
without resorting to exotic explanations such as a ``fifth force" \citep{wen09} or variations on the 
gravitation constant \citep{wen12}: it is enough to suppose that nucleonic 
matter undergoes a transition at high densities  to a phase whose EoS is 
limited only by causality.

On the other hand, comparing Tables~\ref{tab2} and~\ref{tab3} shows that
the effect of quark deconfinement is to reduce the maximum possible 
neutron-star mass, assuming that the speed of sound is limited by $c/\sqrt{3}$. 
We stress, however, that the maximum density of neutron
stars may not be high enough for perturbative QCD to be valid, and that
the EoS of deconfined quarks might well be limited only by causality.

It is interesting to compare our results with those corresponding to the 
assumption of a maximally stiff EoS, i.e., to the assumption that 
the EoS is at the causal limit for all densities above $n_{\rm N}$~\citep{zel62,nau73,rhru74,mal75,bre76,hlc75,
har78,lat90,kal96,kor97,sag12}. The results are given in parentheses in Tables~\ref{tab1} and~\ref{tab2}.
We found, however, that such configurations are thermodynamically unstable. Not surprisingly, 
the maximum mass thus obtained $\sim 3.7M_\odot$ is very high. 
This upper bound has important consequences for identifying compact astrophysical sources.
Taken at face value, it indicates that the soft X-ray transient GRO~J0422$+$32,
whose measured mass is 3.97 $\pm$ 0.95 $M_{\odot}$\citep{gel03}, could be a neutron star. 
On the other hand, the requirement of thermodynamic stability imposes stringent constraints on the maximum 
neutron-star mass. In particular, the maximum mass is found to be reduced by about $1.6-1.7 M_\odot$ 
for the softest of our nucleonic EoSs, as compared to the maximally stiff EoS, as shown in Table~\ref{tab1} and 
Fig.~\ref{fig_MR_exo}. In this case, the identification of GRO~J0422$+$32 as a neutron star is ruled out: it 
must be a black hole.

If the core of neutron stars is made of deconfined quark matter, the speed of 
sound in the maximally stiff EoS 
(obtained by setting $n_{\rm X}=n_{\rm N}$) will be limited by $c/\sqrt{3}$. In this case, the inequality 
$P_{\rm quark}(n)>P_{\rm nuc}(n)$ for $n>n_{\rm N}$, hence also $\mathcal{E}_{\rm quark}(n)>\mathcal{E}_{\rm nuc}(n)$, 
is still satisfied but only in a restricted density range, since the speed of 
sound in nucleonic matter 
generally exceeds $c/\sqrt{3}$ at high enough densities. For this reason, the reduction of the neutron-star 
maximum mass after imposing thermodynamic stability is found to be much less dramatic, amounting to 
$\sim 0.6M_\odot$ at most, as shown in Table~\ref{tab2} and Fig.~\ref{fig_MR_qk}. 

As a matter of fact, the maximum masses of neutron stars without and with quark cores are well approximated by the 
scaling relations \citep{hlc75,har78,wit84}
\beqy 
M_{\rm max} \simeq 4.09 \sqrt{\frac{\mathcal{E}_{\rm nuc}(n_0)}{\mathcal{E}_{\rm exo}(n_{\rm X})}} M_\odot
\eeqy
\beqy
M_{\rm max} \simeq 2.03 \sqrt{\frac{B_0}{B}} M_\odot
\eeqy
respectively, where $B_0=56$ MeV fm$^{-3}$. 

\begin{table*}
\centering
\caption{Global structure of nonrotating neutron stars: maximum mass, 
corresponding radius, corresponding highest baryonic density of 
the nucleonic phase, corresponding lowest baryonic density of 
the exotic phase, and corresponding central baryonic density.}
\label{tab1}
\vspace{.5cm}
\begin{tabular}{|c|ccccc|}
\hline
Force & $M_{\rm max}/M_{\odot}$ & $R$ (km)& $n_{\rm N}$ (fm$^{-3}$)& $n_{\rm X}$ (fm$^{-3}$)&  $n_{\rm cen}$ (fm$^{-3}$)\\
\hline
BSk19 & 2.03 (3.66) & 8.75 (15.62) & 0.20 & 0.63 (0.20) &   1.41 (0.44) \\ 
BSk20 & 2.31 (3.64) & 9.99 (15.54) & 0.20 & 0.49 (0.20) & 1.08 (0.45) \\
BSk21 & 2.42 (3.71) & 10.48 (15.80) & 0.20 & 0.45 (0.20) &  0.99 (0.43) \\ 
\hline
\end{tabular}
\tablefoot{The 
quantities in parentheses refer to the corresponding results for the maximally 
stiff EoS obtained without imposing thermodynamic stability.}
\end{table*}

\begin{table*}
\centering
\caption{Same as Table~\ref{tab1} for hadronic-quark phase transition.}
\label{tab2}
\vspace{.5cm}
\begin{tabular}{|c|cccccc|}
\hline
Force & $M_{\rm max}/M_{\odot}$ & $R$ (km)& $n_{\rm N}$ (fm$^{-3}$)& $n_{\rm X}$ (fm$^{-3}$)&  $n_{\rm P}$ (fm$^{-3}$)& $n_{\rm cen}$ (fm$^{-3}$)\\
\hline
BSk19 & 1.72 (2.26)  & 9.95 (12.93) & 0.20 & 0.34 (0.20) & 0.77 & 1.24 (0.74) \\ 
BSk20 & 1.88 (2.26) & 10.90 (12.93) & 0.20 & 0.29 (0.20) & 0.55 & 1.17 (0.74) \\ 
BSk21 & 2.02 (2.31) & 11.70 (13.24) & 0.20 & 0.25 (0.20) & 0.47  & 0.90 (0.70) \\ 
\hline
\end{tabular}
\end{table*}

\begin{table*}
\centering
\caption{Summary of results from \citet{cfpg11} for maximum neutron-star 
mass without an exotic phase.}
\label{tab3}
\vspace{.5cm}
\begin{tabular}{|c|ccc|}
\hline
Force & $M_{\rm max}/M_{\odot}$ & $R$ (km)& $n_{\rm cen}$ (fm$^{-3}$)\\
\hline
BSk19&    1.86 &  9.13 & 1.45  \\
BSk20&   2.15 &  10.6 & 0.98   \\
BSk21&   2.28 & 11.0 & 0.98   \\
\hline
\end{tabular}
\end{table*}

\section{Conclusions} 
\label{conclusion}

We have investigated the impact of a phase transition in dense matter on the 
structure of neutron stars, considering the stiffest possible EoS constrained by 
i) causality and ii) thermodynamic stability, i.e., the condition that at a given pressure 
the exotic phase should have a lower Gibbs free energy per baryon than 
the nucleonic phase. The latter condition is found to severely limit the maximum mass. 

Even if the phase transition is accompanied by a strong softening of the 
EoS, we find that in the causal limit the maximum possible neutron-star 
mass is always increased above the value determined for a purely nucleonic EoS.
In particular, the softest of our three EoSs, BSk19, will then be able to 
support a neutron star as massive as PSR J1614$-$2230, provided the phase transition 
begins at a density $n_{\rm N}$ as low as $0.2$ fm$^{-3}$. This shows, incidentally, 
that the existence of a two-solar mass neutron star is not necessarily 
incompatible with the soft nuclear-matter EoS that is suggested by the 
measurements of the kaon and pion productions in heavy-ion
collisions \citep{fuchs01,stu01,har06,xiao09}.

On the other hand, the presence of deconfined quarks in dense matter will 
generally lower the maximum mass of neutron stars, assuming the speed of sound 
is limited by $c/\sqrt{3}$, as found by perturbative QCD calculations at zero 
temperature \citep{kur10} and non-pertubative lattice QCD calculations at finite 
temperatures \citep[see e.g.][and references therein]{kar07,bor10}. In this case, 
only the stiffest of our nucleonic EoSs, BSk21,
 will be consistent with the recently measured mass of PSR J1614$-$2230. 
If confirmed, reported observations of significantly more massive neutron stars
with $M>2 M_\odot$ \citep{cla02,fre08,kbk11} will hardly be 
compatible with the presence of quark matter in neutron-star cores \citep[see also][]{lp10} 
unless the sound speed is significantly higher than $c/\sqrt{3}$. 

Considering the current knowledge of dense-nuclear matter properties, it
would be difficult to understand the existence of neutron stars heavier than
$\sim 2.4-2.5 M_\odot$. This upper limit is considerably lower than
estimates that did not impose the constraint of thermodynamical stability.

\begin{acknowledgements}
We are particularly grateful to P. Haensel and J. L. Zdunik for crucial remarks. 
We also had valuable discussions with M. Oertel, S. Reddy, P. Romatschke, and 
A. Sedrakian. The financial support of the FNRS (Belgium), the NSERC 
(Canada) and CompStar (a Research Networking Program of the European Science 
Foundation) is gratefully acknowledged. 
\end{acknowledgements}


\end{document}